\documentclass[%
 reprint,
 amsmath,amssymb,
 aps,
]{revtex4-1}

\usepackage{graphicx}
\usepackage{dcolumn}
\usepackage{bm}
\usepackage{color}
\usepackage{placeins}
\usepackage[caption=false]{subfig}
\usepackage{hyperref}

\begin{document}

\preprint{APS/123-QED}
\title{A simple generalization of Prandtl-Tomlinson model to study nanoscale rolling friction}

\author{Avirup Sircar}
 \affiliation{Department of Mechanical Engineering, Indian Institute of Technology Kharagpur, West Bengal, India - 721302.}
\author{Puneet Kumar Patra}%
 \email{puneet.patra@civil.iitkgp.ac.in}
\affiliation{%
Department of Civil Engineering and Center for Theoretical Studies, Indian Institute of Technology Kharagpur, West Bengal, India - 721302}%

\begin{abstract}
Prandtl-Tomlinson (PT) model has been very successful in explaining nanoscale friction in a variety of situations. However, the simplistic PT model, on account of having a point mass being dragged across a sinusoidal force field, cannot be used for studying rolling friction at nanoscales. In this manuscript, we generalize the PT model as a collection of point particles arranged in a circle of radius $R$. The resulting ``rigid body'' is driven in a composite force field by a moving spring (of stiffness $k$) connected to the center of mass of the rigid body in presence of damping. The force field is a product of the familiar sinusoidal function used in the PT model with a parametrically controlled ($\lambda$) exponentially varying function that is dependent on the vertical coordinates of the particles. Our generalized model degenerates to the standard PT model if $R \ll 1$ and $\lambda \to 0$. With $R \sim 1$ and $\lambda \to 0$, the model undergoes a transition from sticky dynamics to smooth dynamics as $k$ is increased to a critical value. The analytical expression agrees well with the simulation results. Similar analytical expressions have been derived for $ \lambda \neq 0$ as well. In this scenario, the sticky dynamics is experienced in both $x$ and $y$ directions, and our numerical results agree with the analytical solution for $x$ direction. The dynamics, investigated numerically for the general case of $R \sim 1$ and $\lambda \neq 0$, reveals several interesting aspects of nanoscale tribology including the regimes where energy dissipation due to friction is minimum. Further the results from our proposed model are in qualitative agreement with those from MD simulations as well. We believe that the simplicity of our model along with its similarity to the PT model may make it a popular tool for analyzing complicated nanotribological regimes. 
\end{abstract}

\maketitle

\section{Introduction}
Friction is a multiscale phenomenon that occurs when two bodies in contact move relative to each other. It is thought to transform the ordered kinetic energy of the bodies into disordered thermal energy \cite{wang2015energy}. Controlling friction has been a topical research area for well over several decades and the long historical research into friction may be summed up in three simple laws \cite{schwarz2016exploring}: (i) frictional force, $F_f$, that must be overcome to initiate motion is proportional to the normal force, (ii) $F_f$ is independent of the contact area of the two surfaces in contact, and (iii) kinetic friction is independent of the sliding velocity. However, recent research indicates that these ``laws'' \textit{may} not hold true for atomic scale systems \cite{mo2009friction}. With the advent of experimental techniques such as friction force microscope, probing friction at atomic scales has revealed interesting phenomena, such as superlubricity \cite{yang2013observation}, thermolubricity \cite{jinesh2008thermolubricity}, etc., that cannot be explained through the three laws. Thus, we see that the fundamental mechanisms involving friction still require investigations. 

One of the most promising models used for understanding nanoscale tribology is the Prandtl-Tomlinson (PT) model \cite{prandtl1904flussigkeitsbewegung,tomlinson1929cvi}. It comprises a point particle of mass, $m$, being dragged in a sinusoidal force field by a spring of stiffness, $k$, whose other end moves with a constant speed $v$. As the spring elongates, the point particle experiences three forces -- (i) the spring force, (ii) the sinusoidal force, and (iii) the damping force -- the interplay of which causes the particle to move as per following equation:
\begin{equation}
m\ddot{x} = -k \left(x - vt \right) + \dfrac{2 \pi V_0}{a} \sin \left( \dfrac{2 \pi x}{a} \right) - \xi \dot{x}
\label{eq:eq1}
\end{equation}
Here, $V_0$ denotes the amplitude of the potential field, $a$ represents the periodicity of the potential, and $\xi$ represents the damping constant. The PT model agrees with the fundamental properties of friction \cite{popov2010prandtl}: (i) a minimum driving force is required to cause the motion of the particle, which may be interpreted as the static friction force, and (ii) the particle in motion can continue its motion even in presence of a driving force smaller than static friction force owing to its inertia. 

The PT model is a simplification of complicated atomic scale systems such as an atomic force microscope (AFM) moving over a substrate \cite{muser2011velocity}. The sinusoidal force field is a qualitative approximation for the AFM tip - substrate interaction while the spring force is the simplification of the elastic interactions between the AFM's tip and its base. Despite its simplicity, the PT model can explain the stick-slip motion of an AFM \cite{johnson1998stick}. Since the motion of the PT model shows parametric dependence on the dimensionless quantity $\alpha = 4\pi^2V_0/ka^2$, one observes multiple different dynamical regimes. For example, in absence of damping, a smooth motion of the particle is obtained with $\alpha < 1$, while with $\alpha > 1$ the dynamics is discontinuous and a stick-slip motion is observed \cite{medyanik2006predictions}. The value $\alpha=1$ represents the transition from smooth sliding to discontinuous slipping by one lattice site (single-slip regime). In presence of damping (underdamped motion), one may observe multiple stick-slip dynamics for $\alpha > 1$ and $\xi<\sqrt{(V_0/m)}4\pi/a$ \cite{medyanik2006predictions}. Further, the PT model has been used to qualitatively explain tribological mechanisms in a variety of cases: motion over flat surfaces \cite{schwarz2016exploring}, temperature \cite{jansen2010temperature} and velocity dependence of frictional forces \cite{muser2011velocity,gnecco2000velocity}, etc. 

However, being a reduced order model, the traditional PT model cannot explain friction in a variety of situations: (i) in nanoscale systems where the contact between the two nanosurfaces is spread over hundreds (and may be more) of different atomic sites, (ii) in systems where structural effects, like commensurability, play an important role, and (iii) in nanoscale systems undergoing rolling motion. While the first two issues have been tackled through the Frenkel-Kontorova (FK) model, where the standard PT model is generalized to include multiple particles that are coupled elastically with each other and a moving spring \cite{vanossi2013colloquium}, the generalization of PT model to understand rolling friction and its difference from sliding friction \cite{vanossi2013colloquium} is yet to be addressed. This problem has practical relevance as several nanoscale objects such as carbon nanotubes (CNTs) \cite{falvo1999nanometre}, nanoscale bearings comprising double walled CNTs \cite{zhang2004atomistic}, $C_{60}$ molecules \cite{qian2001mechanics}, etc., have been observed to display both rolling and sliding. In certain situations, rolling is favoured over sliding, while in others the reverse is true. For example, through experiments involving the motion of CNTs on a graphitic surface, Buldum and Lu \cite{buldum1999atomic} argued that rolling is favoured over sliding since the energy barrier for sliding is higher. On the other hand, using similar experiments on incommensurate systems Falvo et. al \cite{falvo1999nanometre} observed that in energy dissipation in nanoscale rolling is higher than that in sliding motion. Further, the PT model cannot account for the geometry of the two interacting surfaces \cite{schall2000molecular} and the interactive forces between them \cite{heo2007effect}, which significantly alter the nature of the relative motion (sliding or rolling). 

In this manuscript, we attempt to generalize the PT model to incorporate rolling motion. This is achieved by replacing the point mass of the PT model with $N$ point masses which lie on a circle (of radius $R$) and are connected with each other through Hookean springs. The resulting ring is driven in a damped environment by a spring whose one end is connected to the center of mass (CoM) of the ring and other end is pulled with a constant velocity. The interaction with the substrate is modelled through a force field that varies with the $x$ and $y$ coordinates of the constituent particles of the ring. We shown that in the limit of $R \ll 1$ and asymmetry in force $\to 0$, our model reduces to the traditional PT model. As $R$ and asymmetry in force increases, we observe a plethora of rich dynamics. Our results qualitatively agree with those of molecular dynamics simulations as well. The manuscript is organized as follows: the next section details the proposed generalization of the PT model. The analytical and numerical results are presented next, following which a qualitative comparison with molecular dynamics simulations is presented.

\section{Model and Methodology}
In principle, only bodies having \textit{finite} dimensions can exhibit rolling motion. This is also true in nanoscale where objects with spherical and cylindrical geometries, such as buckyballs and CNTs, have been found to exhibit rolling motion. Taking inspiration from this fact, we generalize the PT model. 
\begin{figure}[b]
\includegraphics[scale=0.5]{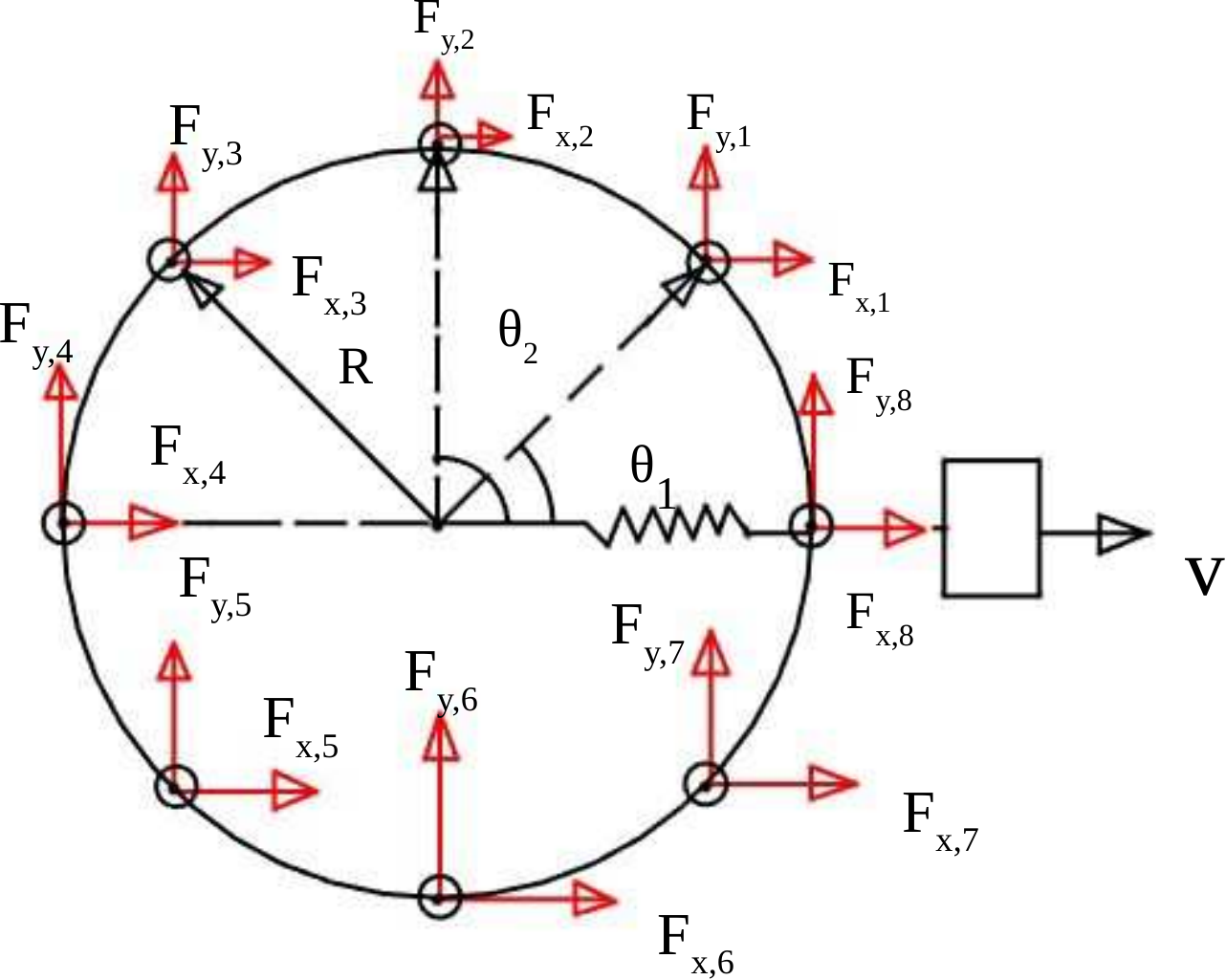}
\caption{\label{fig:fig1}The proposed generalization of the PT model: $N$ particles are arranged on a circle of radius $R$. A particle $i$ makes an angle $\theta_i$ with the $x$ axis. A spring of stiffness $k$ is attached to the CoM of the ring. The other end of the spring is pulled with a constant velocity $v$. Each particle is connected to every other particle through Hookean springs of stiffness $k_{wh}$. The entire system is dragged in a position (both $x$ and $y$) dependent force field in presence of viscous damping. Rolling is facilitated by choosing a potential such that the forces decrease as one goes from bottommost particle to the topmost particle.}
\end{figure}

\subsection{The Model}
The pictorial representation of the proposed model is shown in figure \ref{fig:fig1}. The system comprises $N$ particles, each of mass $m_i$, arranged on a circular ring of radius $R$. The initial coordinates of the $i^{th}$ particle are given by:
\begin{equation}
\begin{array}{rcl}
x_i^0 = R \cos(\theta_i), & & y_i^0 = R \sin (\theta_i) + R, \\
\end{array}
\label{eq:eq2}
\end{equation}
where, $\theta_i$ is the angle with respect to the $x$ axis measured anticlockwise. Each particle of the ring is subjected to the action of four different forces, which we detail next. 

\begin{enumerate}
\item The first set of forces arises because of springs that connect each particle with all other particles. Assuming all these springs have the same stiffness $k_{wh}$, one can write the $x$ and $y$ components of the force on the $i^{th}$ particle as:
\begin{equation}
\begin{array}{rcl}
F^{wh}_{x,i} & = & -k_{wh} \sum\limits_{j \neq i} \dfrac{(r_{ij} - r^{eq}_{ij})(x_i - x_j)}{r_{ij}}, \\
F^{wh}_{y,i} & = & -k_{wh} \sum\limits_{j \neq i} \dfrac{(r_{ij} - r^{eq}_{ij})(y_i - y_j)}{r_{ij}}. \\
\end{array}
\label{eq:eq3}
\end{equation}
Here, $x_k$ and $y_k$ are the instantaneous coordinates of the $k^{th}$ particle and $r_{ij}= \sqrt{(x_i - x_j)^2 + (y_i - y_j)^2}$ is the instantaneous distance between the particles $i$ and $j$. The equilibrium distance between the particles is given by $r^{eq}_{ij}$ and is calculated from the initial geometry. One can make the ring rigid as well as flexible depending on the choice of $k_{wh}$: with $k_{wh} \gg 1$, the ring becomes nearly rigid while for $k_{wh} \sim 1$ one obtains a flexible ring.

\item The next set of forces arises due to a spring of stiffness $k$ whose one end is attached to the CoM of the $N$-particle ring. The other end of the spring is attached to a fictitious mass that moves with a constant speed $v$ along the $x$ axis so that its $y$ coordinate remains fixed at $R$. After a time $t$, the instantaneous magnitude of the net spring force experienced by the ring is given by:
\begin{equation}
F^{sp} = -k \sqrt{(x_c - vt)^2 + (y_c-R)^2 },
\label{eq:eq4}
\end{equation}
where, $x_c = \sum x_i / N$ and $y_c = \sum y_i / N$ denote the instantaneous coordinates of the CoM.  The $x$ and $y$ components of $F^{sp}$ can be written as:
\begin{equation}
\begin{array}{rcl}
F^{sp}_x & = & -k (x_c - vt), \\
F^{sp}_y & = & -k  (y_c - R). 
\label{eq:eq5}
\end{array}
\end{equation}
The net force is assumed to be distributed equally amongst the $N$ particles, so that the spring force on any particle $i$ is:
\begin{equation}
\begin{array}{rcl}
F^{sp}_{x,i} & = & \dfrac{F^{sp}_x}{N} =  \dfrac{-k (x_c - vt)}{N}, \\
F^{sp}_{y,i} & = & \dfrac{F^{sp}_y}{N} = \dfrac{-k (y_c - R)}{N}.
\label{eq:eq6}
\end{array}
\end{equation}

\item The third set of forces occurs because of the viscous medium in which the ring is dragged. With damping constant as $\xi^\prime$, they are dependent on the velocities of the individual particles. Thus, the damping forces on a particle $i$ are given by:
\begin{equation}
\begin{array}{rcl}
F^{d}_{x,i} & = & \xi^\prime \dot{x_i}, \\
F^{d}_{y,i} & = & \xi^\prime \dot{y_i}, 
\label{eq:eq7}
\end{array}
\end{equation}
The net damping forces experienced by the ring may be obtained by summing over the index $i$: $F^d_x = \xi^\prime N \dot{x}_c$ and $F^d_y = \xi^\prime N \dot{y}_c$, where $\dot{x}_c$ and $\dot{y}_c$ indicate the velocity of the CoM. A quick comparison with equation (\ref{eq:eq1}) yields $N \xi^\prime = \xi \implies \xi^\prime = \xi/N$ for maintaining parity with the traditional PT model i.e. the effective damping constant for each particle decreases by a factor $N$ when compared with the PT model.

\item The last set of forces arises because of the potential field in which the ring is pulled. In the traditional PT model, the particle is pulled in a potential field given by: $V = V_0 \cos (2 \pi x / a)$. Let us begin with the same position dependent potential field for our model. The $x$ direction force on the $i^{th}$ particle is given by:
\begin{equation}
\begin{array}{rcl}
F^V_{x,i} & = \dfrac{2 \pi V^\prime_0 }{a} \sin (2 \pi x_i / a)
\nonumber
\end{array}
\end{equation}
The reason for choosing $V_0^\prime $ instead of $V_0$ will be explained later.
However, because of symmetry, no sustained torque can be obtained here. This can be exemplified with a ring of $N=4$. Let the initial $x$ coordinates of the four particles be such that $x_1 = x_4 = -x_2 = -x_3$. The torque of the forces $F^V_{x,i}$ about the CoM sum to 0. 

In our model, a non-zero torque can only be obtained if the forces on two particles, having the same $x$ coordinate, are different. We propose a modified potential field:
\begin{equation}
\begin{array}{rcl}
V_{mod} & = \sum\limits_{i=1}^N\left[ \exp \left( \dfrac{-\lambda y_i^2}{R} \right) \times V_0^\prime \cos \left( \dfrac{2 \pi x_i}{a} \right) \right],
\label{eq:eq8}
\end{array}
\end{equation}
to bring such an asymmetry in the force profile. Here, $\lambda$ is a scaling constant to control the strength of the exponential potential along the $y$ direction. $V_{mod}$ is a more realistic approximation than the traditional PT potential since now we have a situation where the force along the $y$ direction decreases as one goes further from the substrate. Corresponding to $V_{mod}$, the forces on the particles are:
\begin{equation}
\begin{array}{rcl}
F^V_{x,i} & =  \exp \left( \dfrac{-\lambda y_i^2}{R} \right) \times \dfrac{2 \pi V_0^\prime }{a} \sin \left( \dfrac{2 \pi x_i}{a} \right) \\
F^V_{y,i} & = \dfrac{2\lambda y_i}{R} \exp \left( \dfrac{-\lambda y_i^2}{R} \right) \times V_0^\prime \cos \left( \dfrac{2 \pi x_i}{a} \right)
\label{eq:eq9}
\end{array}
\end{equation}
 In the limit $R \to 0, \lambda \to 0$, we have $x_i = x_c$ so that the net $x$ direction force experienced by the ring in this case is: $F^V_x = \dfrac{2 \pi V^\prime_0 N}{a} \sin (2 \pi x_c / a)$. The parity between the proposed model and PT is ensured if $V^\prime_0 N = V_0 \implies V^\prime_0 = V_0/N$. 
\end{enumerate}

Thus, the equations of motion of any particle, $i$, of the ring under the four aforementioned forces are: 
\begin{widetext}
\begin{equation}
\begin{array}{rcl}
m_i \ddot{x}_i & = & -k_{wh} \sum\limits_{j \neq i} \dfrac{(r_{ij} - r^{eq}_{ij})(x_i - x_j)}{r_{ij}} - \dfrac{k (x_c - vt)}{N} - \dfrac{\xi \dot{x_i}}{N} + \exp \left( \dfrac{-\lambda y_i^2}{R} \right) \times \dfrac{2 \pi V_0 }{aN} \sin \left( \dfrac{2 \pi x_i}{a} \right) \\
m_i \ddot{y}_i & = & -k_{wh} \sum\limits_{j \neq i} \dfrac{(r_{ij} - r^{eq}_{ij})(y_i - y_j)}{r_{ij}} - \dfrac{ k (y_c - R)}{N } - \dfrac{\xi \dot{y_i}}{N} + \dfrac{2\lambda y_i}{R} \exp \left( \dfrac{-\lambda y_i^2}{R} \right) \times \dfrac{V_0}{N} \cos \left( \dfrac{2 \pi x_i}{a} \right) \\
\label{eq:eq10}
\end{array}
\end{equation}
\end{widetext}
Assuming $m_i = m_0$ for all particles, the CoM of the ring evolves according to the following equations of motion:
\begin{equation}
\begin{array}{rcl}
 Nm_0\ddot{x}_c & = & -k (x_c - vt) - \xi \dot{x}_c \\
& & + \sum\limits_{i=1}^N\dfrac{2 \pi V_0}{aN} \sin \left( \dfrac{2 \pi x_i}{a} \right) \exp \left( \dfrac{-\lambda y_i^2}{R}\right) \\
Nm_0\ddot{y}_c & = & -k (y_c - R) - \xi \dot{y}_c \\
& & + \sum\limits_{i=1}^N \dfrac{2\lambda y_i}{R} \exp \left( \dfrac{-\lambda y_i^2}{R} \right) \times \dfrac{V_0}{N} \cos \left( \dfrac{2 \pi x_i}{a} \right)\\
\label{eq:eq11}
\end{array}
\end{equation}

\section{Results}
The equations of motion (\ref{eq:eq10}) are solved using $4^{th}$ order Runge-Kutta method for 100 million timesteps where the incremental time step is $10^{-4}$ units. We vary the parameters $k, \xi$ and $\lambda$ independently while keeping $k_{wh} = 1,000,000$ fixed to identify the different regimes of dynamics. The parameter $k_{wh}$ has been  taken high to ensure that the ring behaves rigidly. Our generalization to the PT model can also account for flexible rings, but in the present study, we do not investigate this angle. Note that we do not solve equation \ref{eq:eq11}, but calculate the coordinates of the CoM from the individual coordinates of the different particles. 

The magnitude of friction force is defined in terms of the net force exerted by the moving spring on the ring: $F_f = F^{sp} = -k \sqrt{(x_c - vt)^2 + (y_c-R)^2}$. Unlike the PT model, friction force does not act solely in the $x$ direction, and now has components along both $x$ and $y$ directions.

\subsection{Equivalence with the PT model when $R \ll 1$ and $\lambda \to 0$}
\begin{figure*}
\includegraphics[scale=0.275]{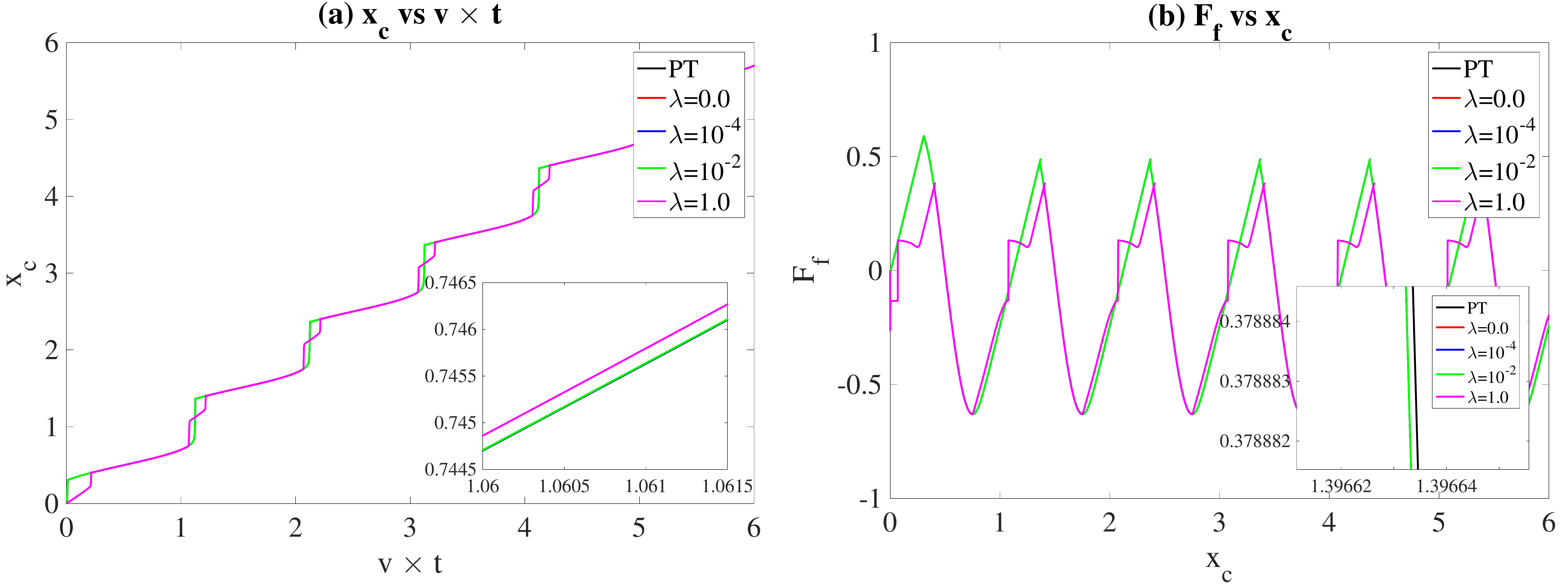}
\caption{Equivalence of the proposed model with the PT model: (a) Position of CoM, $x_c$, of the ring and the PT particle vs. the position of the fictitious mass ($v \times t$), and (b) variation of friction force, $F_f$, with $x_c$. The inset figures show the zoomed in views. For all cases here, the parameters of the models (both PT and proposed) are: $N = 8, m_0 = 1/N$, $R = 0.001$, $k = 2.0$, $\xi = 2\sqrt{2}$, $v = 0.001$, $V_0 = 0.1$ and $a = 1.0$. In the proposed model, $\lambda$, that governs the strength of the interaction in the $y$ direction is varied from $\lambda = 0$ to $1$. As is evident, from the figures, our proposed model is equivalent to the standard PT model when $R \ll 1$ and $\lambda \to 0$. However, with increasing $\lambda$ the deviation from PT model increases.}
\label{fig:fig2}
\end{figure*}

Equation \ref{eq:eq11} simplifies to the following with $\lambda \to 0$ and $R << 1$:
\begin{equation}
\begin{array}{rcl}
Nm_0\ddot{x}_c & = & -k (x_c - vt) - \xi \dot{x}_c + \dfrac{2 \pi V_0}{a} \sin \left( \dfrac{2 \pi x_c}{a} \right) \\
Nm_0\ddot{y}_c & = & -k (y_c - R) - \xi \dot{y}_c 
\label{eq:eq12}
\end{array}
\end{equation}
Under these conditions, since there is no excitation along the $y$ direction, the equations of motion (\ref{eq:eq12}) reduce to PT equations of motion (\ref{eq:eq1}) with $m_0 = 1/N$. 

We now show numerically that the proposed model is equivalent to the standard PT model under these conditions. For this purpose, the equations of motion of the PT model, shown in (\ref{eq:eq1}), are solved numerically with the following parameters: $k = 2.0$, $\xi = 2\sqrt{2}$, $v = 0.001$, $V_0 = 0.1$ and $a = 1.0$. For the proposed model, we choose the same values of the parameters while solving the equation (\ref{eq:eq10}) with the following additions: $N = 8,m_i = 1/N = 1/8$, $R = 0.001$. The parameter, $\lambda$, that governs the strength of the $y$ direction potential in the proposed model is varied from $\lambda = 0$ to $\lambda = 1.0$. Figure \ref{fig:fig2} plots: (a) the variation of $x_c$ and the position of the PT particle with respect to the position of fictitious mass ($v\times t$) and (b) the variation of friction force, $F_f$, for the proposed model and the PT model vs. $x_c$.  It can be seen from the figures that our proposed model gives results in agreement with the PT model under $R \ll 1$ conditions when $\lambda \to 0$. As $\lambda$ increases, the disparity in the dynamics increases. It is interesting to note that for each case investigated here, we do not observe any rolling whatsoever. Rather, the entire ring gets dragged across the substrate. For the case of $\lambda = 1.0$, the CoM displays a large displacement along both $x$ and $y$ directions. The trajectory indicates that the ring is dragged along the substrate following which it smoothly hops and reaches quickly a minimum energy configuration. During these hoppings no rotation is observed.

\begin{figure}[htbp]
\includegraphics[scale=0.275]{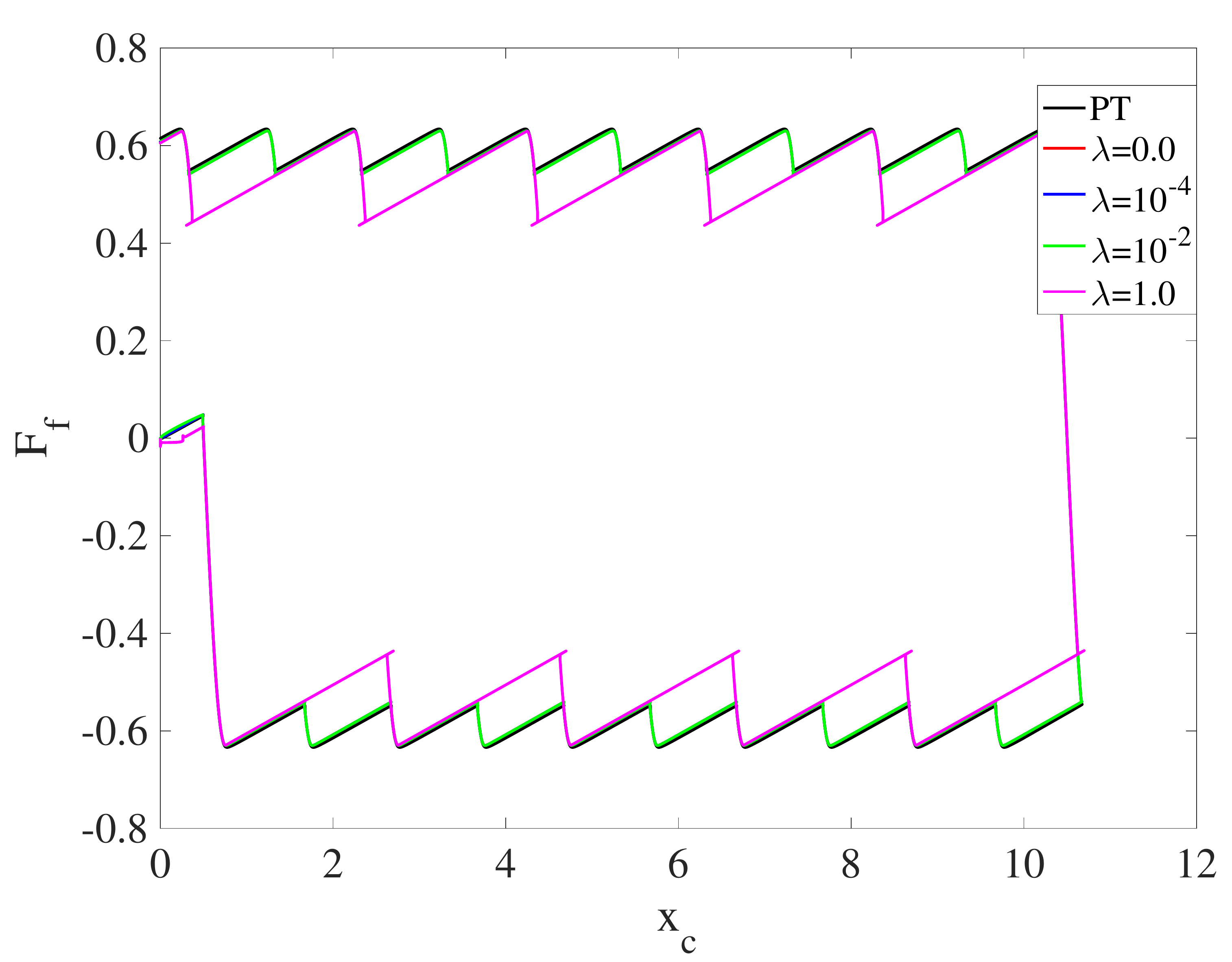}
\caption{This figure shows the equivalence of stick-slip and hysteretic behavior between the proposed model and the PT model when $\lambda \to 0$. All parameters of the remain the same as before except $k$ which now takes a value of 0.10. As $x_c$ exceeds 10.0 units, $v$ is reversed.}
\label{fig:fig3}
\end{figure}

The success of PT model lies in its ability to demonstrate rich dynamics which depends on the choice of the parameters. For example, only by changing $k$ from 2.0 to 0.10 makes the dynamics stick-slip, and if one subsequently reverses $v$, hysteretic behavior is obtained. In a similar way, the model proposed in the present study is able to demonstrate rich dynamics on changing the parameters. Figure (\ref{fig:fig3}) shows the equivalence of the stick-slip and hysteretic behavior between the proposed model and the PT model when $\lambda \to 0$. Like in the previous case, the disparity between the two models increases with increasing $\lambda$. 

A common theme emerges from figures (\ref{fig:fig2}) and (\ref{fig:fig3}) -- in small-radius small-$\lambda$ limit, the proposed model behaves identically with the PT model. The equivalence between the two models makes us believe that both the models will behave similarly in other cases as well, and hence, we do not explore this angle further in the present work.

\subsection{From smooth sliding to sticky dynamics}
\begin{figure}[htbp]
\includegraphics[scale=0.25]{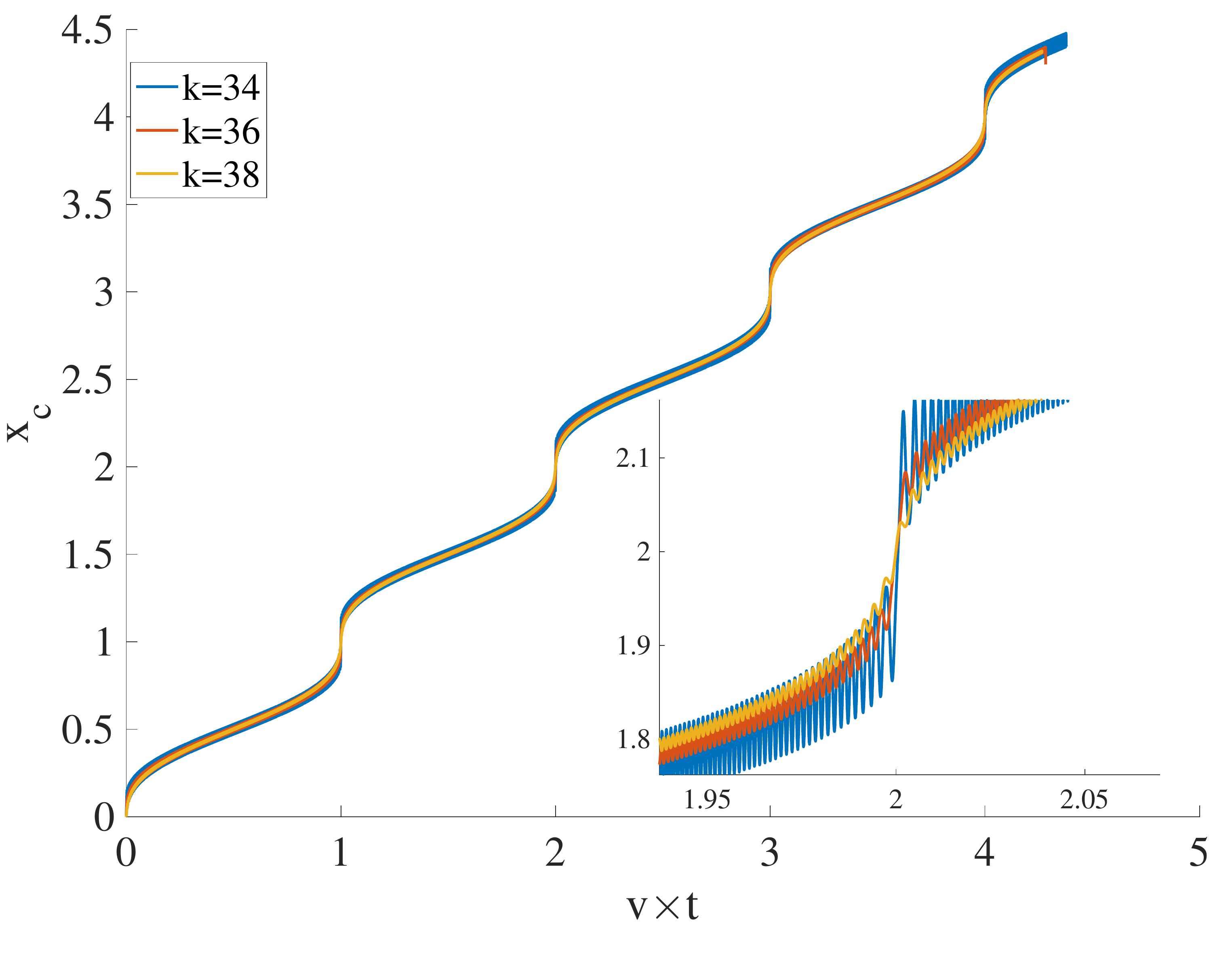}
\caption{Transition from smooth sliding to stick-slip dynamics -- As $k$ is decreased from 38 to 34, the dynamics changes from smooth sliding to stick-slip. The properties of the ring are: $R = 0.10$, $N = 8, m_0 = 1/N$, $a = 1.0$, $V_0 = 1.0$ and $v = 0.001$. The inset figure shows the zoomed in view where the jump is clearly visible at smaller values of $k$. The transition occurs at $k \approx 36$ which agrees with the theoretical prediction.}
\label{fig:fig4}
\end{figure}
In absence of damping with $\lambda \to 0$, the ring reaches an equilibrium state if $F_x = 0$:
\begin{equation}
\begin{array}{rcl}
F_x = 0 & \to & -k (x_c - vt) + \sum\limits_{i=1}^N\dfrac{2 \pi V_0}{aN} \sin \left( \dfrac{2 \pi x_i}{a} \right)  = 0
\label{eq:eq13}
\end{array}
\end{equation}
Without the loss of generality, $x_i$ at any time $t$ may be written as: $x_i = x_c + R \cos \left(\dfrac{2 \pi i}{N} \right)$. Substituting it back in equation (\ref{eq:eq13}) gives:
\begin{equation}
\begin{array}{rcl}
k (x_c - vt)  & = & \sum\limits_{i=1}^N\dfrac{2 \pi V_0}{aN} \sin \left( \dfrac{2 \pi}{a}\left(x_c + R \cos\left(\dfrac{2 \pi i}{N}\right)\right) \right) 
\label{eq:eq14}
\end{array}
\end{equation}
The instability in dynamics occurs when the resulting equilibrium is unstable, i.e. $\dfrac{\partial F_x}{\partial x_c} > 0$, which gives us:
\begin{equation}
\begin{array}{rcl}
k \leq \dfrac{4 \pi^2}{a^2} \dfrac{V_0}{N} \left[ \sum\limits_{i=1}^{N} \cos \left( \dfrac{2 \pi x_c}{a} + \dfrac{2 \pi R}{a} \cos \left( \dfrac{2 \pi i}{N}\right)\right) \right]
\label{eq:eq15}
\end{array}
\end{equation}
Expanding equation (\ref{eq:eq15}) and realizing that $\sum\limits_{i=1}^N \sin \left[ \dfrac{2 \pi R}{a} \cos \left( \dfrac{2 \pi i}{N}\right) \right]= 0$ for even $N$ and rigid ring, equation (\ref{eq:eq15}) gets simplified to:
\begin{equation}
\begin{array}{rcl}
k \leq \dfrac{4 \pi^2}{a^2} \dfrac{V_0}{N}  \cos \left( \dfrac{2 \pi x_c}{a} \right)\left[ \sum\limits_{i=1}^{N} \cos \left( \dfrac{2 \pi R}{a} \cos \left( \dfrac{2 \pi i}{N}\right)\right) \right]\\
\implies k \leq \dfrac{4 \pi^2}{a^2} \dfrac{V_0}{N} \left[ \sum\limits_{i=1}^{N} \cos \left( \dfrac{2 \pi R}{a} \cos \left( \dfrac{2 \pi i}{N}\right)\right) \right]
\label{eq:eq16}
\end{array}
\end{equation}
The transition between smooth motion and sticky motion for the ring occurs when $k = k_{crit} = \dfrac{4 \pi^2}{a^2} \dfrac{V_0}{N} \left[ \sum\limits_{i=1}^{N} \cos \left( \dfrac{2 \pi R}{a} \cos \left( \dfrac{2 \pi i}{N}\right)\right) \right]$. For all values of $k< k_{crit}$ one obtains sticky motion while $k \geq k_{crit}$ results in smooth motion. In the continuum limit of $N \to \infty$, the value of $k_{crit}$ takes the form as:
\begin{equation}
\begin{array}{rcl}
k_{crit} & = & \dfrac{1}{2\pi}\int\limits_{0}^{2\pi}\left(\dfrac{2\pi}{a}\right)^2V_0 \left[ \cos\left(\dfrac{2\pi x_c}{a}+\dfrac{2\pi R}{a}\cos\left(\eta\right)\right)\right]d\eta \\
& = & \dfrac{4 \pi^2}{a^2}V_0\mathbb{J}_0\left(\dfrac{2\pi R}{a}\right),
\end{array}
\label{eq:eq17}
\end{equation}
where, $\mathbb{J}_0$ is the Bessel function of the first kind. It is interesting to note that $k_{crit}$ is independent of $N$ in the continuum limit and only depends on the radius of the ring. We now test the accuracy of the developed formulation using numerical simulations. Consider a ring with $R = 0.10$, $N = 8, m_0 = 1/N$, $a = 1.0$, $V_0 = 1.0$ and $v = 0.001$. The numerical results for the same are shown in figure (\ref{fig:fig4}). As $k$ is decreased from 38 to 34, the dynamics changes from smooth sliding to stick-slip. The inset figure shows the zoomed-in view where the jump is clearly visible at smaller values of $k$. The transition occurs at $k \approx 36$ which agrees with the theoretical prediction of $k_{crit}$ obtained from equation (\ref{eq:eq16}).

We now study the regime where $\lambda \neq 0$ and $R << 1$. In this regime, $x_i \approx x_c$ and $y_i \approx y_c$ so that the forces on the CoM along the $x$ and $y$ directions take the form: 
\begin{equation}
\begin{array}{rcl}
F_x & = & \dfrac{2\pi V_0}{a}\exp\left({-\dfrac{\lambda}{R} y_c^2} \right)\sin\left(\dfrac{2\pi}{a}x_c\right)-k(x_c-vt)\\
F_y & = & \dfrac{2V_0\lambda}{R}\exp \left({-\dfrac{\lambda}{R} y_c^2} \right) \cos\left(\dfrac{2\pi}{a}x_c\right)-k(y_c-R)
\label{eq:eq18}
\end{array}
\end{equation}
As in the previous case, the instability occurs when $\max \left( \dfrac{\partial F_\alpha}{\partial \alpha_c} \right) \geq 0$, where $\alpha = \{x,y\}$. Following conditions emerge as a result:
\begin{equation}
\begin{array}{rcl}
\dfrac{\partial F_x}{\partial x_c} \geq 0 & \implies & k \leq \dfrac{4 \pi^2}{a^2}V_0 \exp \left(-\dfrac{\lambda}{R}y_c^2 \right) \\
& \implies & k \leq  \dfrac{4 \pi^2}{a^2}V_0\\
& \implies & k_{crit} =  \dfrac{4 \pi^2}{a^2}V_0 \\
\dfrac{\partial F_y}{\partial y_c} \geq 0 & \implies & k \leq -\dfrac{4 \lambda^2}{R^2}V_0 y_c \exp \left(-\dfrac{\lambda}{R}y_c^2 \right) \\
& \implies & k \leq \left(\dfrac{2 \lambda}{R}\right)^{3/2}\dfrac{V_0}{\sqrt{e}} \\
& \implies & k_{crit} = \left(\dfrac{2 \lambda}{R}\right)^{3/2}\dfrac{V_0}{\sqrt{e}}
\label{eq:eq19}
\end{array}
\end{equation}
The third inequality of equation (\ref{eq:eq19}) follows from the fact that $\max\left( \exp\left( -\dfrac{\lambda}{R}y_c^2\right)\right) = 1$ when $y_c = 0$. The third inequality of the second row of the above equation follows from: $\max\left(-2\beta y \exp(-\beta y^2) \right) = \sqrt{\dfrac{2 \beta }{e}}$. While the first case denotes the stability criterion for the $x$ direction, the second one is for the $y$ direction. The global dynamics is smooth when $k > k_{crit}^{max}$, where $k_{crit}^{max}$ is the maximum of the two criteria obtained in equation (\ref{eq:eq19}). 

\begin{figure}[htbp]
\includegraphics[scale=0.275]{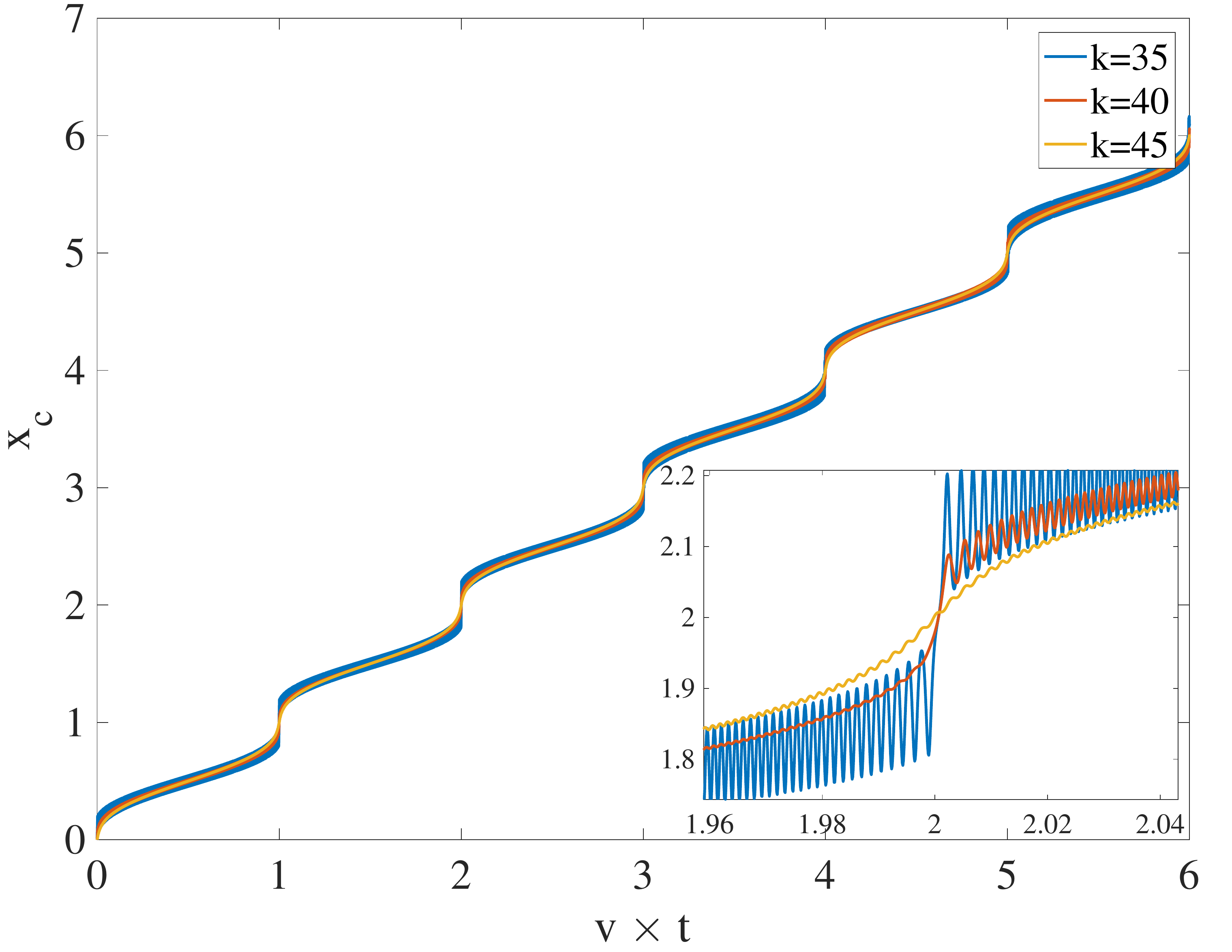}
\caption{Transition from sticky to smooth motion when stability is governed by the $x$ direction criterion: $k_{crit} =  \dfrac{4 \pi^2}{a^2}V_0$. With $\lambda = 0.001$, $R = 0.001$, $N = 8, m_0 = 1/N$, $a = 1.0$, $V_0 = 1.0$ and $v = 0.001$ theoretical value of $k_{crit}\approx 40$. It is evident from the figure that for $k<40$ there is a sudden jump in the $x$ direction which disappears for $k \geq 40$. }
\label{fig:fig5a}
\end{figure}

We now verify the results numerically. Two test cases are designed such that each criterion separately governs the stability: (i) $\lambda = 0.001$, $R = 0.001$, $N = 8$, $a = 1.0$, $V_0 = 1.0$ and $v = 0.001$ where $k_{crit}^{max}\approx 40$ is obtained from $\partial F_x / \partial x_c >0$, and (ii) $\lambda = 0.01(0.02)$ with rest of the details same as in (i) where $k_{crit}^{max} \approx 55(154)$ is obtained from $\partial F_y / \partial x_y >0$. Figure (\ref{fig:fig5a}) shows the transition from sticky to smooth motion when stability is governed by the $x$ direction criterion: $k_{crit}^{max} =  \dfrac{4 \pi^2}{a^2}V_0$. It is evident from the figure that for $k<40$ there is a sudden jump in the $x$ direction which disappears for $k \geq 40$. However, for our second test case, the numerical results (not shown) differ from that of analytical expression. 

In the most generalized case, where $\lambda \neq 0$ and $R > 0$, a closed form solution of $k_{crit}$ is difficult to obtain, and hence, one has to rely completely on simulations. 

\subsection{Dynamics at $\lambda \neq 0$ and $R>0$}
We now study numerically the dynamics under general situation. Separate parametric studies have been performed to understand the dependence of frictional forces on -- (i) increasing the asymmetric nature of forces by changing $\lambda$, (ii) increasing the velocity of pulling, $v$, and (iii) increasing the radius of the ring, $R$. We now present each of these results.

\subsubsection{Parametric Dependence on $\lambda$}
\begin{figure*}[htbp]
\includegraphics[scale=0.335]{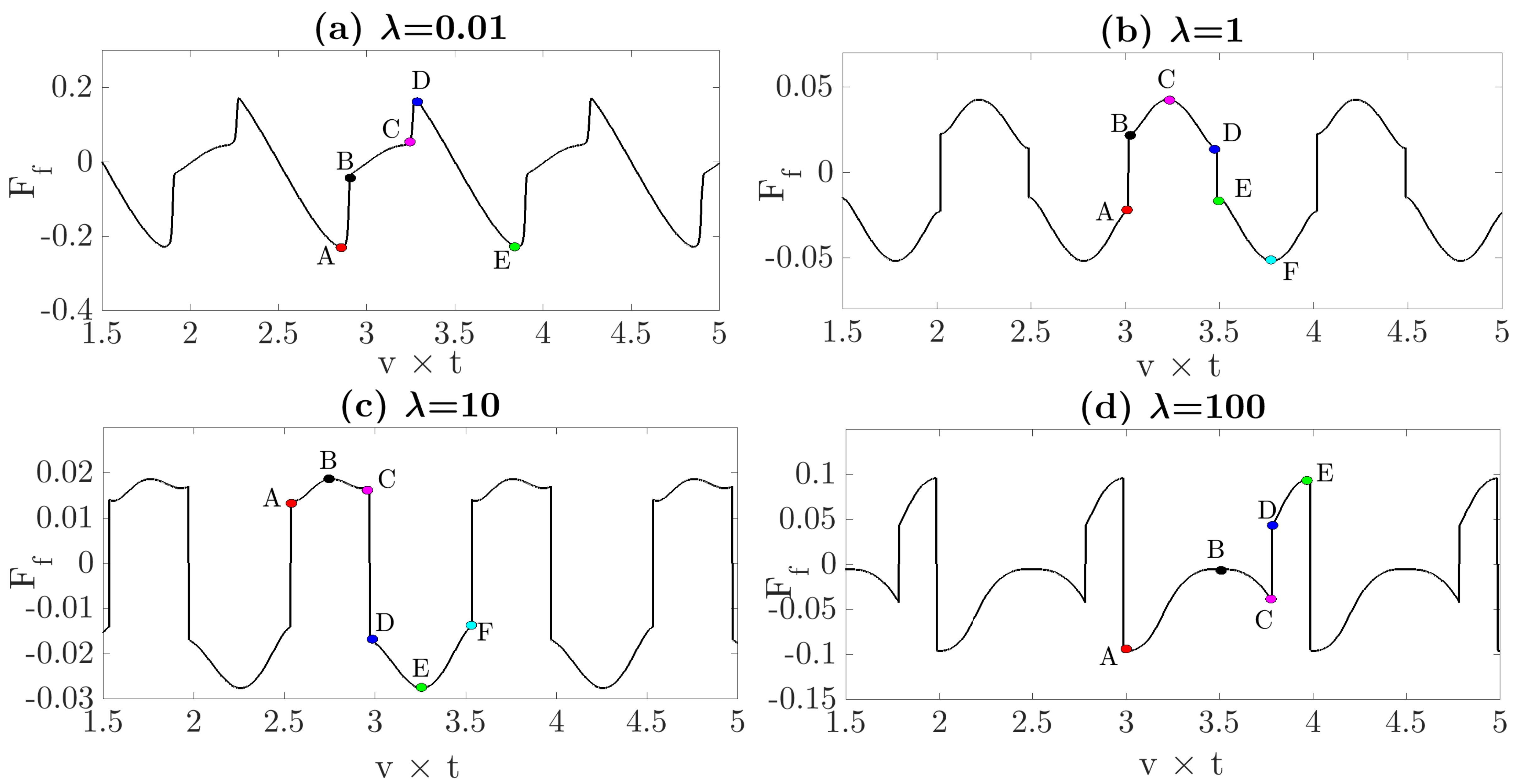}
\caption{Friction vs $v \times t$ for different $\lambda$ values. The properties of the ring are: $N = 8, m_0 = 1/N, R = 1.0, k = 2, \xi = 2 \sqrt{2}, v = 0.001, a = 1.0, V_0 = 0.10$ and $k_{wh} = 10^6$. The ring rolls in each of the cases. For low values of $\lambda$, a combination of dragging-rolling is observed which gradually turns into continuous rolling. Each salient phase has been marked in the figure, which will be described separately.}
\label{fig:fig5}
\end{figure*}

Recent research indicates that friction at nanoscale is influenced significantly by the nature of interactive forces between the two bodies \cite{heo2007effect}. This phenomenon is difficult to capture in the traditional PT model, as the depth of the potential well, $V_0$, is the only term controlling the strength of interaction. In the proposed model, $\lambda$, that determines the asymmetry in the forces is better suited for understanding the influence of interactive forces. The simulations used to show the dependence of frictional forces on the nature of interactive forces involves a ring with the following properties: $N = 8, m_0 = 1/N, R = 1.0, k = 2, \xi = 2 \sqrt{2}, v = 0.001, a = 1.0, V_0 = 0.10$ and $k_{wh} = 10^6$. $\lambda$ is varied from 0.001 to 100 in multiples of 10. With increasing $\lambda$, the torque experienced by the ring increases, which eventually causes the ring to display rich rolling dynamics. The variation of friction is shown in figure (\ref{fig:fig5}) for four values of $\lambda$. 

\begin{figure*}[htbp]
\includegraphics[scale=0.335]{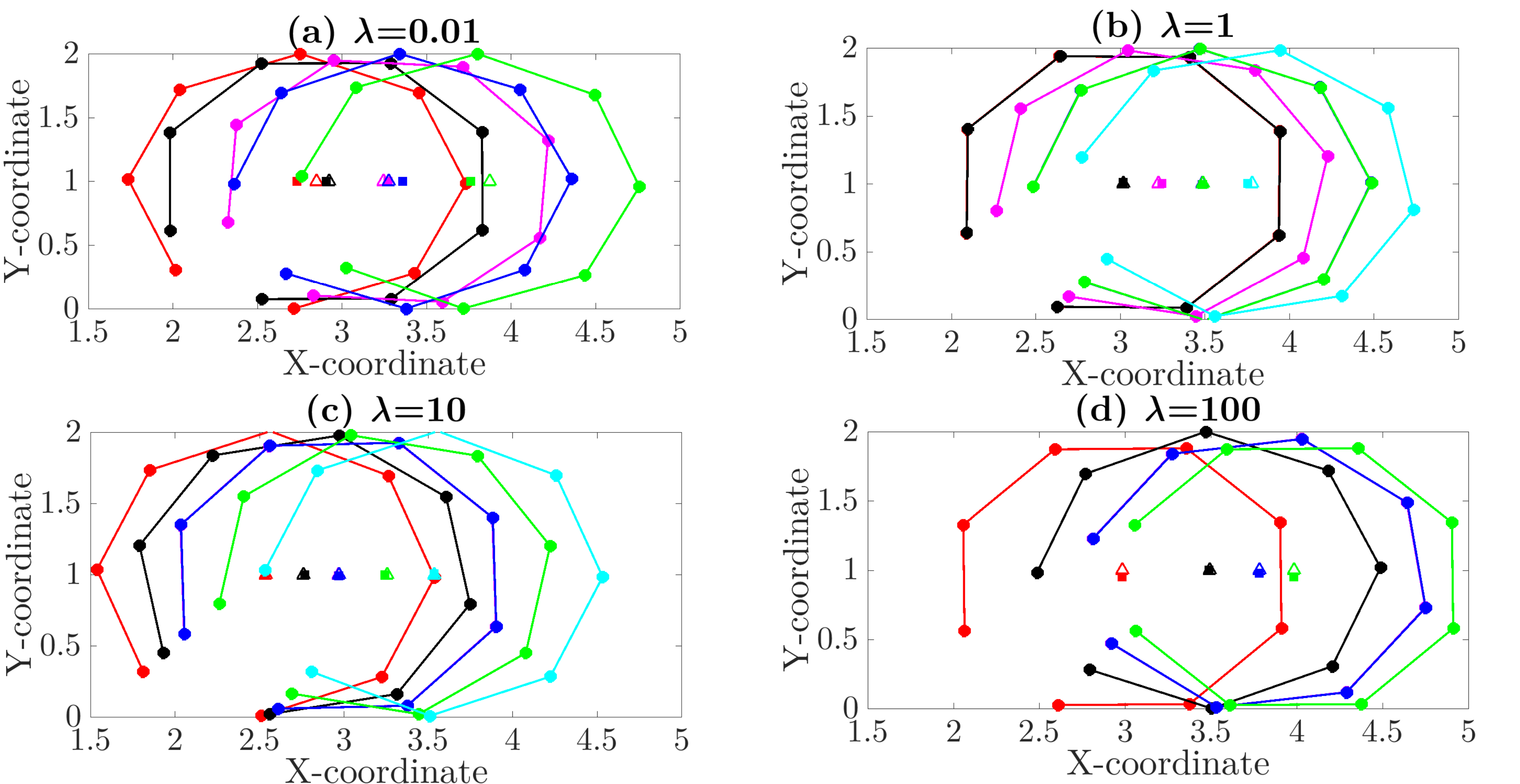}
\caption{Position of the ring corresponding to the salient phases marked in figure (\ref{fig:fig5}). Color code of the rings has been chosen as per the color of the salient phase. The CoM is shown in solid squares while the empty triangles depict the position of the fictitious particle.}
\label{fig:fig6}
\end{figure*}

For $\lambda = 0.01$ a combination of dragging and rolling is observed (see figure \ref{fig:fig6}(a)) -- the ring rolls about the bottommost point during the motion phase A-B and friction quickly changes in this phase, subsequently the ring gets dragged during the motion phase B-C wherein there is a gradual change in friction. Another phase of rolling (C-D) and dragging (D-E) occur next. This pattern keeps getting repeated. As $\lambda$ increases to 1.0, the ring undergoes a combination of dragging and (mostly) rolling motion for the majority of time (see figure \ref{fig:fig6}(b)) -- during the motion phase B-C, C-D and E-F. The sudden change in the direction of frictional force direction during the motion phase A-B and D-E occurs because of the reversal of position of CoM with respect to the fictitious mass. The pattern remains the same as $\lambda$ increases further. It is easy to observe that the slope of friction vs $v \times t$ is an indicator of the dynamics of the ring -- small slope indicates dragging, a relatively larger slope indicates rolling while a slope of $\infty$ indicates reversal of position of CoM with respect to fictitious mass.

\subsubsection{Parametric Dependence on $v$}
\begin{figure*}[htbp]
\includegraphics[scale=0.31]{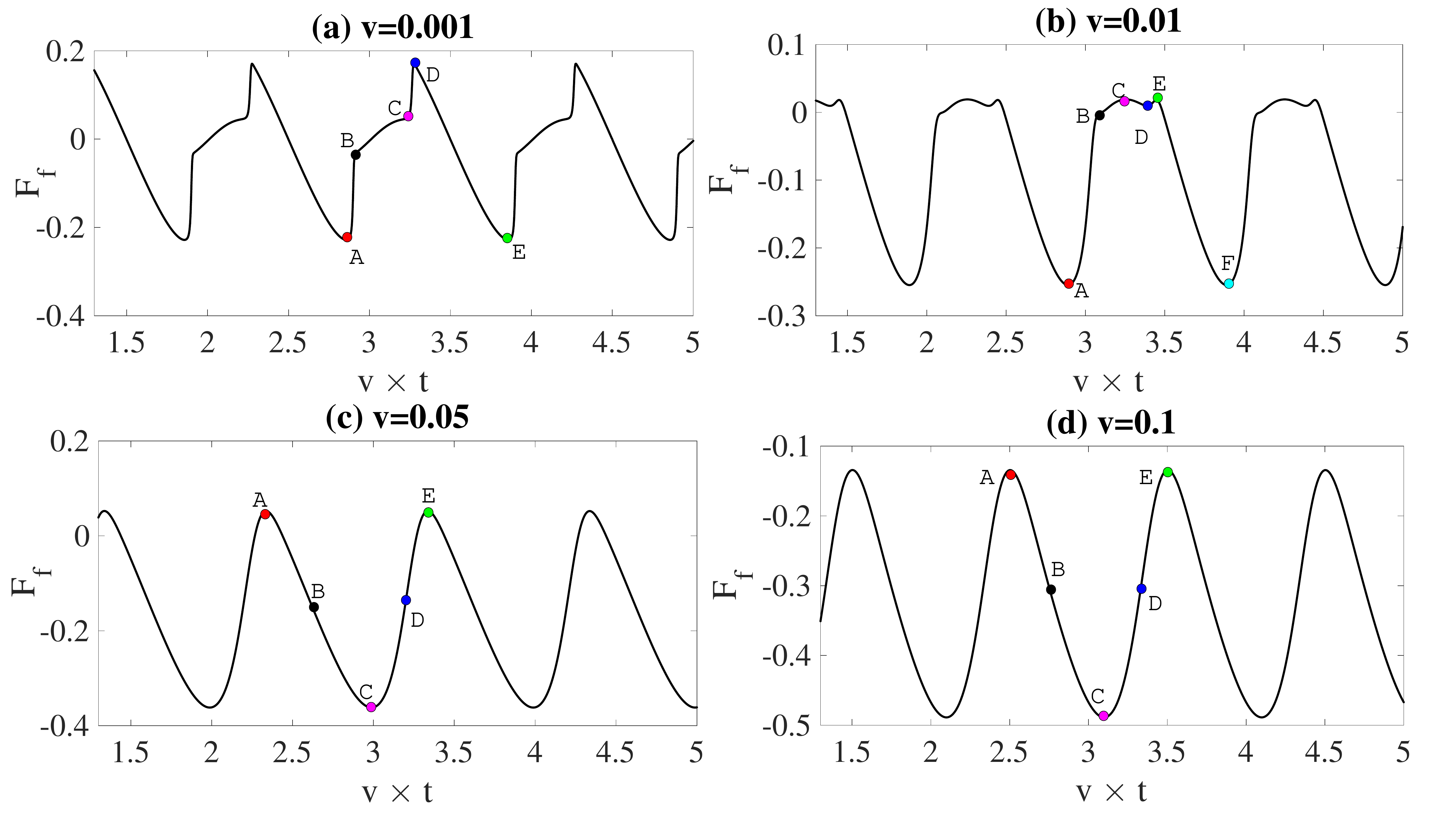}
\caption{Dependence of frictional force on the velocity of pulling the fictitious particle. The properties of the ring are: $N = 8, m_0=1/N, R = 1.0, k = 2, \xi = 2 \sqrt{2}, \lambda = 0.01, a = 1.0, V_0 = 0.10$ and $k_{wh} = 10^6$. Like in PT model, the smooth dragging is associated with a smooth frictional force without any ``kinks''.  On the other hand two different types of kinks can be observed each for $v =0.001,0.01$. The frictional forces pre and post these kinks have a different slope, and like before, one can associate these kinks with rolling and dragging motion of the ring.}
\label{fig:fig7}
\end{figure*}

\begin{figure*}[htbp]
\includegraphics[scale=0.324]{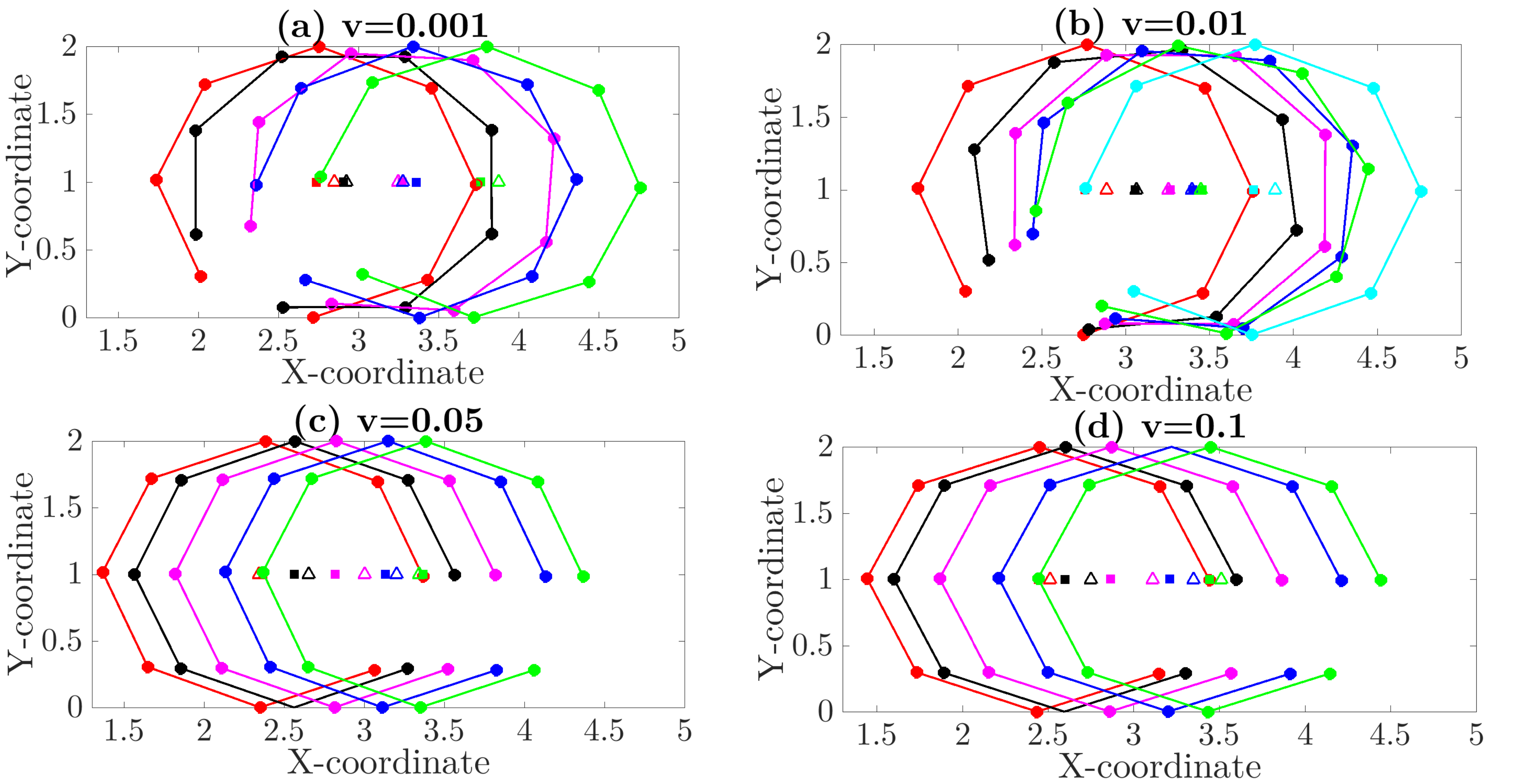}
\caption{Position of ring corresponding to the salient features observed in figure \ref{fig:fig7}. With increasing $v$, the ring dynamics changes from a combination of dragging-rolling to smooth dragging.}
\label{fig:fig8}
\end{figure*}

We now study the dependence of frictional force on the velocity, $v$, with which the fictitious particle is pulled. Keeping all other properties of the system fixed: $N = 8, m_0=1/N, R = 1.0, k = 2, \xi = 2 \sqrt{2}, \lambda = 0.01, a = 1.0, V_0 = 0.10$ and $k_{wh} = 10^6$, $v$ is increased systematically from 0.0005 to 0.10. The resulting friction vs. $v \times t$ plots are shown in figure \ref{fig:fig7}. As $v$ increases, (i) the magnitude of frictional force increases, and (ii) the dynamics of the ring changes from a combination of dragging-rolling to smooth dragging. The latter is evident from the trajectory of a single particle (not shown) as well as the positions of the ring shown in figure (\ref{fig:fig8}) -- one can observe that for $v = 0.05$ and 0.10, the the ring has a translation motion, indicating smooth dragging while the particles of the ring drag and roll at smaller velocities. Corresponding to smooth dragging, the friction profile is smooth as well, without any observable kinks. However, at smaller velocities, where the particle drags as well as rolls, there are ``kinks'' in the friction profile. These kinks are characterized by a sudden change in the slope of the frictional force. Like in the previous case, here as well, we observe that the slope of friction vs $v \times t$ curve is more when rolling occurs than sliding. 

\subsubsection{Parametric Dependence on $R$}
\begin{figure*}[htbp]
\includegraphics[scale=0.35]{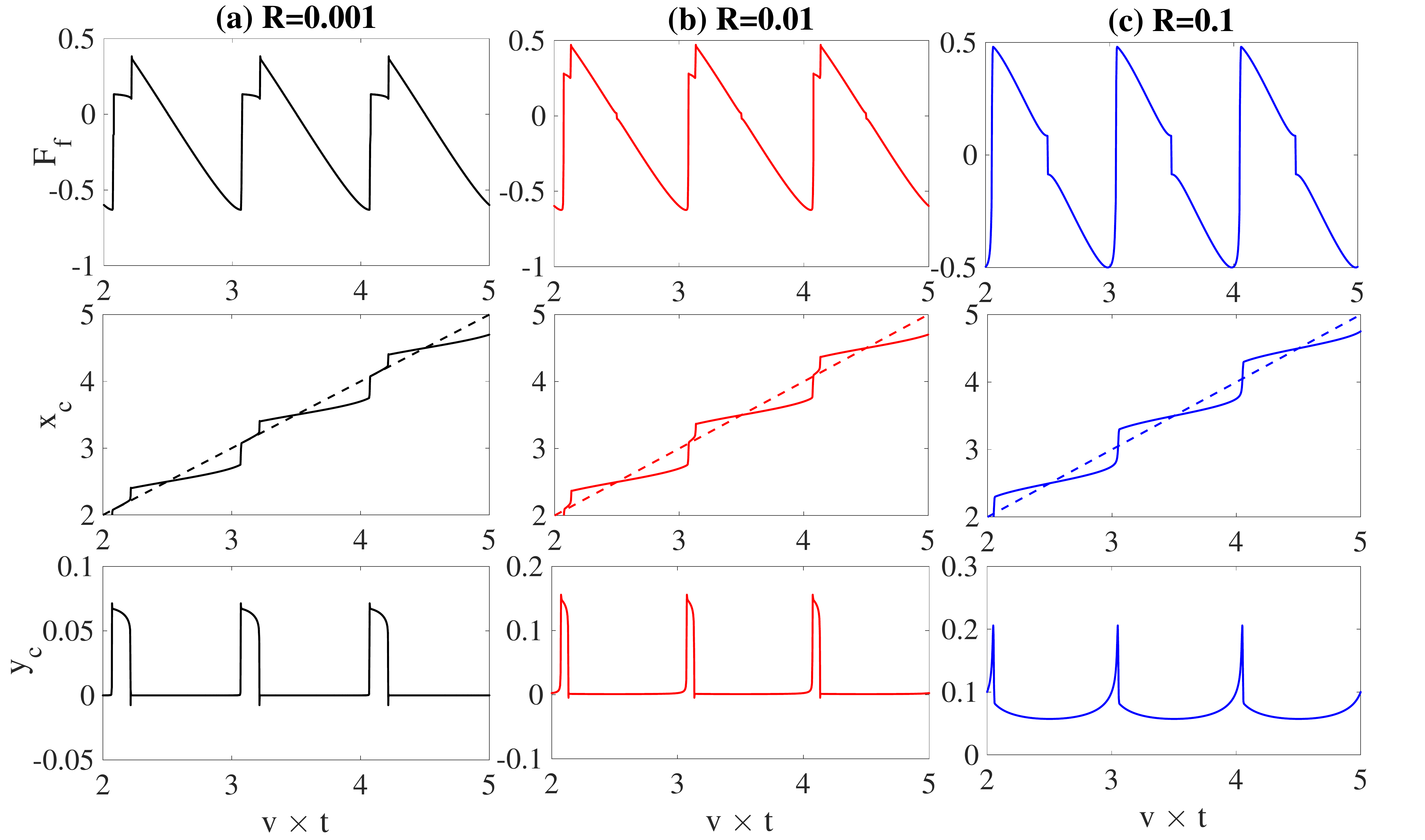}
\caption{Dependence of $F_f$ on radius, $R$, of the ring keeping all other parameters constant. The top row shows the variation of $F_f$ with $v \times t$ as the radius changes: (a) $R$ = 0.001, (b) $R$ = 0.01 and (c) $R$ = 0.1. The solid line of middle row shows the variation of $x_c$ as a function of $v \times t$ for the three cases while the bottom row shows the variation of $y_c$ as a function of $v \times t$. The dashed lines of middle row corresponds to straight lines with unit slope. The locations where the dashed line intersects the solid line indicate the positions where the fictitious particle is at the same position as the CoM of the ring. The dynamics for each case has an instability around the location of substrate where $y_c$ suddenly increases. The ring, in all the cases shown here, mostly gets dragged followed by hopping over the substrate particles. At larger radii (not shown), the ring undergoes a combination of rolling-dragging motion. }
\label{fig:fig9}
\end{figure*}

We now study the parametric dependence on $R$, keeping all other properties of the ring constant: $N = 8, m_0=1/N,v = 0.001, k = 2, \xi = 2 \sqrt{2}, \lambda = 1.0, a = 1.0, V_0 = 0.10$ and $k_{wh} = 10^6$. The variation of $F_f$ with $R$ is shown in the top row of figure \ref{fig:fig9}. The middle row shows the variation of $x_c$ as a function of $v \times t$ for the three cases while the bottom row shows the variation of $y_c$ as a function of $v \times t$. The dynamics for each case has an instability around the location of substrate where $y_c$ suddenly increases. At very small radius, the CoM of ring mostly remains around $y = 0$. However, at larger radius, $R = 0.1$, the CoM of the ring never reaches $y = 0$. 

The frictional force profile suggests that the rings at small radius mostly get dragged, except when they are near the substrate positions. At these locations, the rings undergo a hopping motion without any rotation. A hopping event can be identified by the sudden increase in the $y_c$, and is marked by sudden changes in the frictional force profile. Whether the hopping motion is continuous or interrupted with dragging is determined by the radius: whereas for $R=0.1$ the hopping is continuous for $R = 0.001$ a hopping event is followed with intermittent dragging. This is evident from the frictional profile for $R= 0.001, 0.01$, where $F_f$ increases abruptly only to show a slight decrease followed by another abrupt increase. At $R=0.10$, a different feature emerges more prominently -- the locations where the fictitious particle overtakes the CoM of the ring causes sudden reversal in the sign of the frictional force (see around $v \times t = 2.5, 3.5$ in the top row of figure \ref{fig:fig9}). The ring shows a combination of dragging-rolling for larger radii (not shown here).

\subsection{What is energetically more favorable -- Rolling or Dragging?}
We now try to answer the question -- is rolling favored over dragging? In order to do so, we quantify $\langle W_d \rangle$, the average work done by the frictional forces per unit distance moved by the fictitious particle, where,
\begin{equation}
\langle W_d \rangle = \dfrac{ v \int\limits_{\tau = 0}^{t}F^{sp}_x d\tau}{v \int\limits_{\tau=0}^t d\tau},
\end{equation}
as $\lambda$ increases from 0 to a large value. It is obvious that at $\lambda = 0$, regardless of the other variables, the ring will (mostly) undergo dragging. With $\lambda$ set to a non-zero value, the forces become asymmetric, enabling the ring to undergo rotation. The results for $\langle W_d \rangle$ as $\lambda$ increases from 0 in steps of 10 are tabulated in table \ref{tab:tab1}. As the motion transitions from dragging $\to$ dragging-rolling $\to$ rolling, the magnitude of the work done first decreases and subsequently increases. It attains a minimum value for $\lambda = 1$, which suggests that the amount of energy expended in moving the ring is minimum in this case. We believe that for every combination of ring properties, there exists a $\lambda$ that minimizes the work done by friction. From energy perspective, it seems that rolling is favorable over dragging, however, extensive simulations with our model requires to be performed to come up with a definitive answer. 

\begin{table}
   \centering
   \begin{tabular}{|p{2.2cm}|p{2.2cm}|p{2.8cm}|}
 \hline
 $\lambda$ & $\langle W_d \rangle$ & Motion type\\ [4pt]
 \hline
   0.0 &$-7.059 \times 10^{-3}$  &Dragging \\
   \hline
  0.001 &$-4.445 \times 10^{-3}$ &Dragging-Rolling  \\
   \hline
    0.01 &$-3.203 \times 10^{-3}$ &Dragging-Rolling  \\
   \hline
  0.1 & $-9.154 \times 10^{-4}$ &Dragging-Rolling  \\
    \hline
  1 & $-4.921 \times 10^{-4}$ &Rolling  \\
    \hline
  10 &$-5.771 \times 10^{-4}$ &Rolling  \\
    \hline
  100 & $-5.875 \times 10^{-4}$ &Rolling  \\
   \hline
     \end{tabular}
   \caption{Average work done per unit distance moved by the fictitious particle, $\langle W_d \rangle$ as $\lambda$ increases. All other properties of the ring are kept constant: $N = 8, m_0 = 1/N, R = 1.0, k = 2, \xi = 2 \sqrt{2}, v = 0.001, a = 1.0, V_0 = 0.10$ and $k_{wh} = 10^6$.}
   \label{tab:tab1}
\end{table}

\section{Comparison with Molecular Dynamics}
In this section, we compare qualitatively the results from our model with those from molecular dynamics (MD) simulations. We first describe the methodology adopted for the MD simulations and then describe the results.

\subsection{MD Simulation Methodology}
\begin{figure}[t]
\centering
\includegraphics[scale=0.340]{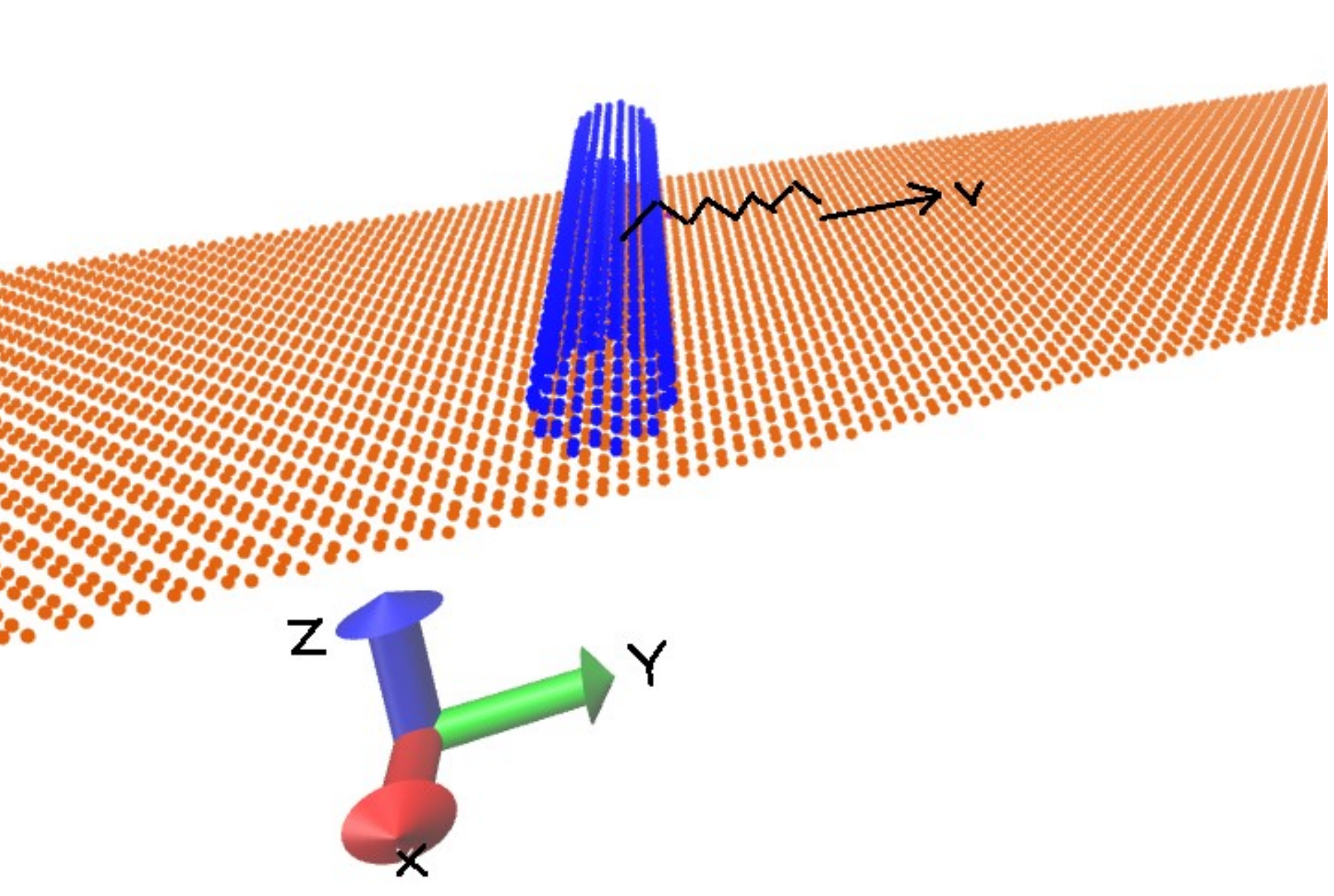}
\caption{\label{fig:fig11} Pictorial representation of the problem studied through molecular dynamics simulation. The graphene substrate is shown in brown dots and the SWCNT in blue dots. The SWCNT moves due to the action of a spring whose one end is connected to the CoM of the SWCNT. The free end of the spring is pulled with a constant velocity of $v$.}
\end{figure}

The pictorial representation of the problem is shown in figure (\ref{fig:fig11}). A single walled (10,0) zigzag Carbon nanotube (SWCNT) of length 100 \AA (shown in blue dots) moves over a fixed graphene substrate of dimensions 100 \AA $\times$ 500 \AA (shown in brown dots). A spring of stiffness of 1 eV/\AA$^2$ (shown in black line) is connected to the CoM of the SWCNT. The free end of the spring is pulled with a velocity $v$ in the $y$ direction. The pull force exerted on the SWCNT by the spring is resisted by the van der Waals forces due to the interaction between SWCNT and graphene. 

A three body Tersoff-like potential \cite{tersoff1988new} has been adopted for modelling the interaction between the C-C atoms of both SWCNT and graphene. The reason for choosing Tersoff-like potential is due to its wide usage in the molecular dynamics community for investigating nanoscale friction in CNTs  \cite{lindsay2010optimized,suekane2008static,chen2008temperature,barreiro2008subnanometer,servantie2006rotational}. Tersoff potential is represented by the following equation:
\begin{eqnarray}
E=\sum_i{E_i}=\frac{1}{2}\sum_{i\neq{j}}\sum_j{\phi(r_{ij}}),\nonumber\\
\phi(r_{ij})=f_c(r_{ij})[f_R(r_{ij})+b_{ij}f_A(r_{ij})],
\label{eq:ab}
\end{eqnarray}
where, $E$ denotes the total potential energy of the system, $E_i$ the potential energy of the $i^{th}$ atom, and $\phi$ the potential energy between the $i^{th}$ and $j^{th}$ atoms. The other variables of equation (\ref{eq:ab}) signify the following: $r_{ij}$ represents the distance between the $i^{th}$ and the $j^{th}$ atoms, $b_{ij}$ is the bond order function, $f_c$ the cutoff function that ensures nearest-neighbor interaction, $f_R$ the repulsive pair potential, and $f_A$ the attractive pair potential. The mathematical forms of these individual functions are given below:
\begin{eqnarray}
f_c(r_{ij}) = \left\{
     \begin{array}{lr}
       \text{1}&\forall  {r}_{ij}<{P}_{ij}\nonumber\\
       \frac{1}{2}-\frac{1}{2}\text{sin}(\frac{\pi}{2}\frac{r_{ij}-R_{ij}}{D_{ij}})& \forall{P}_{ij}<{r}_{ij}<{Q}_{ij}\nonumber \\
       \text{0}&\forall{r}_{ij}>{Q}_{ij}\nonumber\\
     \end{array}
   \right.
     \end{eqnarray} 
\begin{eqnarray} 
f_{R}(r_{ij})=Ae^{-\lambda_1 r_{ij}},f_A(r_{ij})=-Be^{-\lambda_2 r_{ij}},\nonumber\\  
b_{ij}=(1+\beta^n\zeta_{ij}^n)^{-\frac{1}{2n}},\nonumber\\
\zeta_{ij}=\sum_{k\neq i,j}f_C(r_{ik})g(\theta_{ijk})\text{exp}[\lambda_3^{3}(r_{ij}-r_{ik})^3],\nonumber\\
g(\theta_{ijk})=1+c^2/d^2-c^2/[d^2+(h-\text{cos} \theta_{ijk})^2)],                       
\end{eqnarray}
Here, ${P}_{ij}={R}_{ij}-{D}_{ij}  ,  {Q}_{ij}={R}_{ij}+{D}_{ij}$. The cutoff function is continuous and goes from 1 to 0 smoothly as the distance varies from $P_{ij}$ to $Q_{ij}$. $R_{ij}$ is chosen so as to include only the first-neighbor shell for most problems of interest. $\theta_{ijk}$ is the angle between the bonds ${ij}$ and ${ik}$. Depending upon the system being simulated, the parameters in the equation take different values. In the present work parameters proposed by Lindsay and Broido \cite{lindsay2010optimized} have been adopted.  

The cross-interaction between the C-C atoms of SWCNT and graphene is modelled through the two-body Lennard-Jones (LJ) potential:
\begin{eqnarray}
\phi_{LJ}(r_{ij}) = 4\epsilon
\left[\left(\frac{\sigma}{r_{ij}}\right)^{12}-\left(\frac{\sigma}{r_{ij}}\right)^6\right],
\nonumber
\end{eqnarray}
Here, $\epsilon$ is the depth of the potential well and equals 0.002411 eV. $\sigma$ denotes the distance between any two atoms such that $\phi_{LJ}(r_{ij}) = 0$, and equals 3.4 \AA.

The free-to-use LAMMPS software \cite{plimpton1995fast} has been used for MD simulations. The system comprising graphene and SWCNT is equilibrated at a constant temperature environment of 1K through a Nos\'e-Hoover chain thermostat \cite{martyna1992nose,hoover2015ergodic,patra2015deterministic} for 200,000 time steps, where each time step corresponds to 1 fs. In order to maintain equivalence with our model, the graphene atoms are kept fixed during these equilibration as well as subsequent runs. After the system gets equilibrated, the thermostat is removed and the free end of the spring is moved with a constant velocity, $v$. The SWCNT atoms, as a result of experiencing a pull force due to spring, start to move as well. In order to make an equivalent comparison with our model, each atom of SWCNT is subjected to a viscous damping force, where the damping constant is chosen as 0.001 eV-ps/\AA$^2$. Friction is defined as the force that resists the motion of SWCNT in the $y$ direction i.e. it is the $y$ component of the force of interaction between the SWCNT and graphene.


\subsection{Comparison of MD and proposed model}

\begin{figure*}[htbp]
\centering
\includegraphics[scale=0.425]{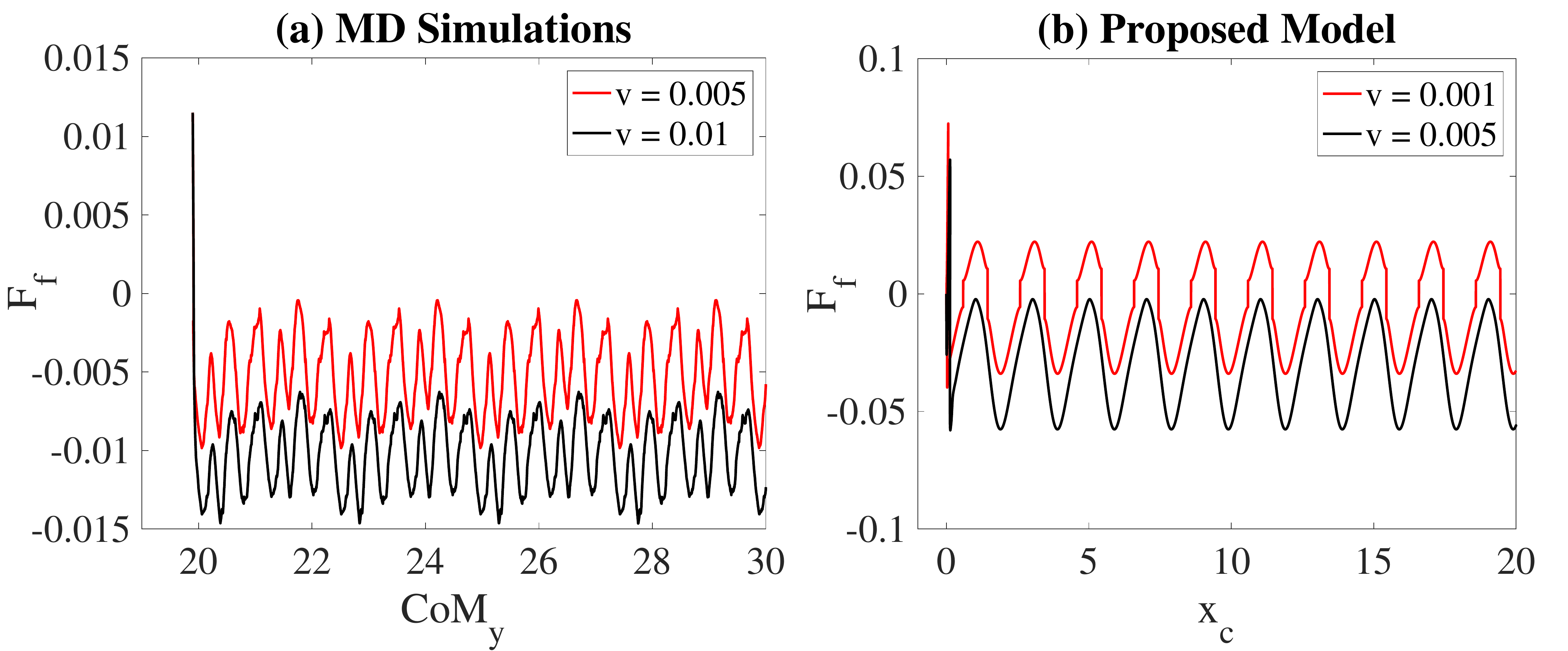}
\caption{\label{fig:fig12} A qualitative comparison of results obtained from MD simulations and those obtained from our proposed model: (a) frictional force vs. $y$ coordinate of CoM of the SWCNT for two different pulling velocities as obtained from MD simulations, and (b) frictional force vs. $x_c$, the CoM of the ring, as obtained from our proposed model. Both the models have a good qualitative agreement -- a combination of dragging and rolling is seen in both the models, frictional force shows an initial quick decrease, and the magnitude of averaged frictional force is more when the driving velocity is higher.}
\end{figure*}

We now compare the results obtained from MD simulations and those obtained from our proposed model. Note that this is \textit{only} a qualitative comparison. A quantitative comparison may be possible by arduously tuning the different parameters of our model. The parameters of the proposed model are chosen as follows: $N=10$ as each ring of a (10,0) SWCNT comprises 10 Carbon atoms, $R=3.1910$ keeping $a=1.0$ such that $a/R$ ratio remains the same as that of graphene-SWCNT system, $k=2.0$, $V=0.1$, $\lambda = 1.0$ and $\eta = 2\sqrt{2}$.

The results obtained from MD simulations and our proposed model are shown in figure (\ref{fig:fig12}) for two separate velocity of pulling: $v = 0.005$ \AA/ps and $0.01$ \AA/ps for MD simulations and $v = 0.001$ and $0.005$ for our model. In both MD and our proposed model, we observe a combination of dragging and rolling motion. The similarity between the two models is quite evident from the frictional force profiles: (i) the frictional force shows a quick initial decrease which may be thought of as static frictional force, and (ii) the magnitude of frictional force, in an averaged sense, is more when the driving velocity is higher.

\section{Conclusions}
In this work, we have tried to remove the shortcomings associated with the Prandtl-Tomlinson (PT) model -- its inability to account for (i) rolling dynamics and rolling friction, (ii) geometrical shape of the objects, and (iii) complicated interactive forces between two nanoscale objects. This is achieved by replacing the point mass of the PT model with a collection of particles that form a ring. Each particle of the ring interacts with every other particle through a Hookean spring, and is subjected to a position dependent potential field along with viscous damping. The center of mass of the ring is connected to a spring whose other end is pulled with a constant velocity $v$. In order to achieve rotation in this system, an asymmetric force profile is needed. We accomplish this by means of a composite potential field that varies along both the $x$ (sinusoidal) and $y$ (exponential) directions. The resulting asymmetry can be controlled parametrically ($\lambda$), so that this model can be effectively used for studying both sliding/dragging dynamics as well as rolling dynamics.

In the limit of $\lambda \to 0$ and $R \ll 1$, the proposed model behaves identically with the traditional PT model -- the frictional force profile as well as the displacement time history are nearly the same and so is the hysteretic behavior. In this regime, like the PT model, our proposed model transitions from smooth sliding/dragging to sticky (dragging/rolling) dynamics as the stiffness, $k$, of the pulling spring increases. An analytical expression of critical stiffness, $k_{crit}$ that causes similar transition when $R \sim 1$ has been obtained and verified through numerical simulation. Similar expressions have also been established for non-zero $\lambda$, where the sticky dynamics may occur in both $x$ and $y$ directions. However, our numerical simulations agree with analytical $k_{crit}$ \textit{only} for the $x$ direction. We are yet to find the reasons behind disagreement for the transition in $y$ direction.

The rich dynamics exhibited by the ring suggests that one can obtain a variety of different frictional force profile and displacement time-history depending upon the controlling parameter. Interestingly, certain combinations of the parameters yield small energy dissipated per unit distance, and they invariably correspond to the situation where the ring executes rolling dynamics. We, therefore, argue that at atomistic scales, rolling is favored over sliding only if the controlling parameters support it. This raises the possibility of engineering super-lubricity by fine-tuning the controlling parameters without altering the physical characteristics of the nano-objects. A quick comparison of our model with that of molecular dynamics simulations shows that the results from both the approaches agree qualitatively. 

The proposed model is very general and may be used to study a variety of different problems, for example, the behavior of flexible circular nano-objects, effect of normal reaction on tribological characteristics of nano-objects, the effect of non-circular geometry and its role in enhancing frictional dissipation, etc. Additionally, by appropriately varying $\lambda$, one can also study the role of interactive forces between two nanoscale objects. However, there remains a scope of further development of the model, for example, the position of substrate particles being variable, the effect of mobile substrate particles, etc. 

We believe that our model overcomes some of the biggest shortcomings of the PT model while still maintaining its familiarity and simplicity. 

\section{Acknowledgment}
Support for the research provided in part by Indian Institute of Technology Kharagpur under the grant DNI is gratefully acknowledged. 

\bibliographystyle{vancouver}
\bibliography{references}
\end{document}